# Research on Personalized Financial Product Recommendation by Integrating Large Language Models and Graph Neural Networks

Yushang Zhao*

McKelvey School of Engineering, Washington University in St. Louis, St. Louis, USA, *Corresponding author: yushangzhao@wustl.edu

Yike Peng

Graduate School of Arts and Sciences,Columbia University, New York, NY, USA, yp2425@columbia.edu

Dannier Li

School of Computing, University of Nebraska - Lincoln,Lincoln, NE, USA, dannierli@outlook.com

Yuxin Yang

Fu Foundation School of Engineering and Applied Science, Columbia University, New York, NY, USA, yy3277@columbia.edu

Chengrui Zhou

Fu Foundation School of Engineering and Applied Science, Columbia University, New York, NY, USA, zhou.chengrui@columbia.edu

Jing Dong

Fu Foundation School of Engineering and Applied Science ,Columbia University, New York, NY, USA, jd3768@columbia.edu

**Abstrac**t: With the rapid growth of fintech, personalized financial product recommendations have become increasingly important. Traditional methods like collaborative filtering or content-based models often fail to capture users' latent preferences and complex relationships. We propose a hybrid framework integrating large language models (LLMs) and graph neural networks (GNNs). A pre-trained LLM encodes text data (e.g., user reviews) into rich feature vectors, while a heterogeneous user–product graph models interactions and social ties. Through a tailored message-passing mechanism, text and graph information are fused within the GNN to jointly optimize embeddings. Experiments on public and real-world financial datasets show our model outperforms standalone LLM or GNN in accuracy, recall, and NDCG, with strong interpretability. This work offers new insights for personalized financial recommendations and cross-modal fusion in broader recommendation tasks.

## 1 INTRODUCTION

The rise of financial technology has driven the use of personalized recommendation in financial services. Traditional methods capture explicit preferences but struggle with heterogeneous data and implicit user needs. LLMs extract deep interests from text, while GNNs model user-product and social interactions. This paper proposes an LLM-GNN hybrid framework: LLMs encode text into feature vectors, and GNNs refine these with graph-structured data via collaborative message passing. Experiments on public and real-world datasets show notable improvements in precision, recall, and NDCG, with enhanced interpretability.

## 2 LITERATURE REVIEW

### 2.1 Applications of Large Language Models (LLMs) in Recommendation Systems

LLM-based recommendation methods fall into two types: discriminative and generative. Discriminative models treat recommendations as classification or regression tasks, using either fine-tuning or prompt tuning. Fine-tuning yields strong performance but is resource-intensive and less adaptable, while prompt tuning is efficient but may lack accuracy and stability [1-3].

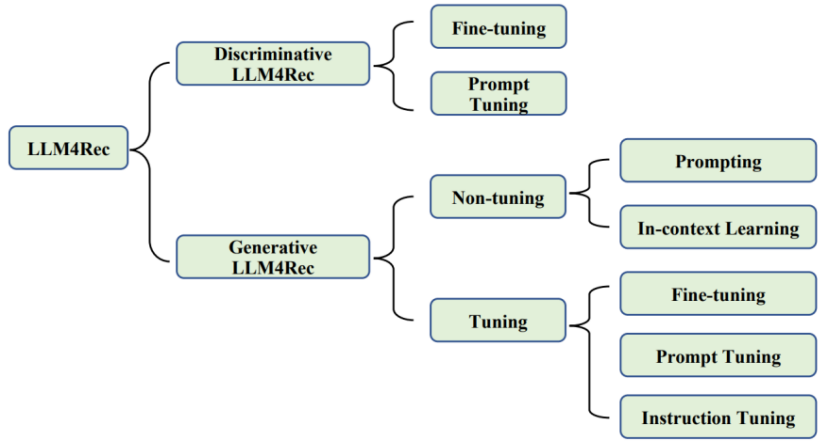

Figure 1: Framework of LLM4Rec method classification and tuning strategy

Generative LLM4Rec approaches transform recommendation into natural language generation using prompting or in-context learning[4-6]. While they offer adaptability and interpretability with minimal labeled data, their performance relies heavily on prompt design. Tuning strategies like instruction tuning improve precision, but challenges remain in achieving the robustness of discriminative models[7].

Discriminative LLM4Rec models offer high accuracy but face high costs and overfitting risks, while generative models provide flexibility and interpretability at the expense of consistency. Future research should explore hybrid approaches—combining multi-task learning, prompt engineering, and graph-based methods— to harness the strengths of both paradigms [8-10].

### 2.2 Advances in Graph Neural Networks (GNNs) for Personalized Recommendation

Based on the traditional graph neural network, this study proposes a personalized recommendation model with multi-level semantic aggregation and interest-aware fusion as shown in Figure 2.

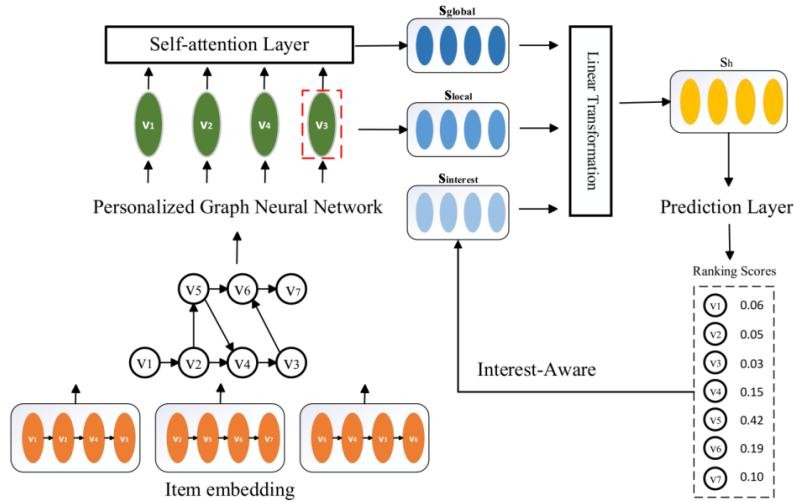

Figure 2 Personalized graph neural network recommendation framework

In the item embedding phase, user behavior is decomposed into multiple interaction paths (e.g., V1–V2–V4–V3), which are vectorized via a shared embedding layer to capture sequential dependencies. These subsequences are fused into a unified heterogeneous interaction graph, representing diverse relationships among users and items [11-13]. A self-attention GNN module computes both global (graph-wide) and local (neighbor-based) aggregation features in parallel, balancing overall coherence and local precision [14-17]. To enhance responsiveness to recent behaviors, an interest-aware branch extracts features from the latest interaction (e.g., $V_3$), which are combined with global and local signals via a linear transformation for unified feature fusion [18-20]. The final prediction layer maps the fused high-order node representation ShS_hSh to ranking scores for candidate items, optimized end-to-end using cross-entropy or ranking loss. Experiments on financial recommendation datasets show improvements in Hit Rate, NDCG, and MRR, while the interest branch enhances interpretability and trust. This framework integrates multi-path context, self-attention, and short-term interest modeling, offering a recommendation system that balances accuracy, efficiency, and explainability for financial services [21-23].

## 3 METHODOLOGY

### 3.1 Model Architecture Design Combining LLM and GNN

The hybrid model in this study operates across three collaborative stages: GNN-driven LLM, GNN-LLM co-driven, and LLM-driven GNN. As shown in Figure 3, the first stage, GNN-driven LLM, employs a graph neural network to perform structural learning on the user-item interaction graph, extracting high-order topological features and neighbor influences to generate embeddings that capture social relationships and behavioral patterns. These graph-based embeddings are then passed into a large language model (LLM) via an interface module to enrich the semantic representations of users and items, enhancing the expressiveness of textual features [24-27].

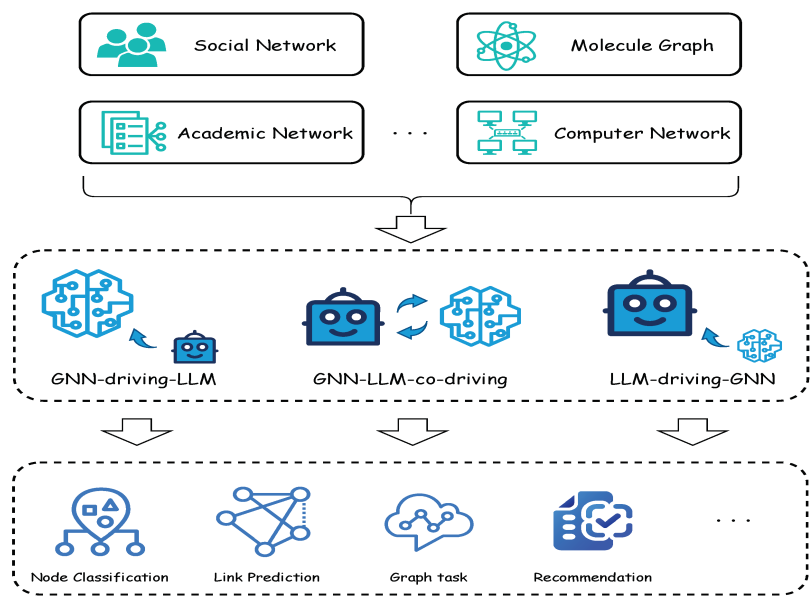

Figure 3 Overall framework of LLM-GNN collaborative driven recommender system

In the GNN-LLM collaborative stage, the model iteratively optimizes graph structures and text features together: LLM embeddings enrich GNN node features, while GNN-aggregated structural information is fed back into the LLM for deep cross-modal fusion. This enhances both models' ability to capture relational and semantic information, overcoming single-model limitations on heterogeneous data. In the LLM-driven GNN stage, LLM-generated semantic vectors from text (e.g., user comments or reports) supplement node features or predict graph structures, enabling the GNN to better perform tasks like node classification and edge prediction [28-30].

The advantages of this architecture are: (1) enabling end-to-end learning of diverse information types (behavior sequences, social relationships, text descriptions) within a unified framework; (2) allowing shared representations between graph tasks (e.g., node classification, link prediction) and recommendation tasks to balance generalization and task-specific accuracy; and (3) improving model interpretability and stability through alternating driving and collaborative feedback. In practical financial product recommendation scenarios, this three-stage pipeline can be flexibly adapted based on data availability and business needs, offering an efficient and deployable solution for personalized recommendations in multi-source heterogeneous environments [30-33].

### 3.2 Feature Representation, Interaction Mechanisms, and Algorithmic Model

In the personalized recommendation of financial products, different data sources have complementary values: the text attributes can reflect the implicit preferences of users and the characteristics of items, and the interaction graph structure reveals the complex relationship between user-item and item-item[34-37]. To this end, based on the above macro architecture as shown in Figure 4, this study designs two parallel and complementary feature stream processing paths, and introduces a pseudo-label mechanism and a multi-objective optimization strategy to achieve deep fusion and co-optimization in end-to-end training.

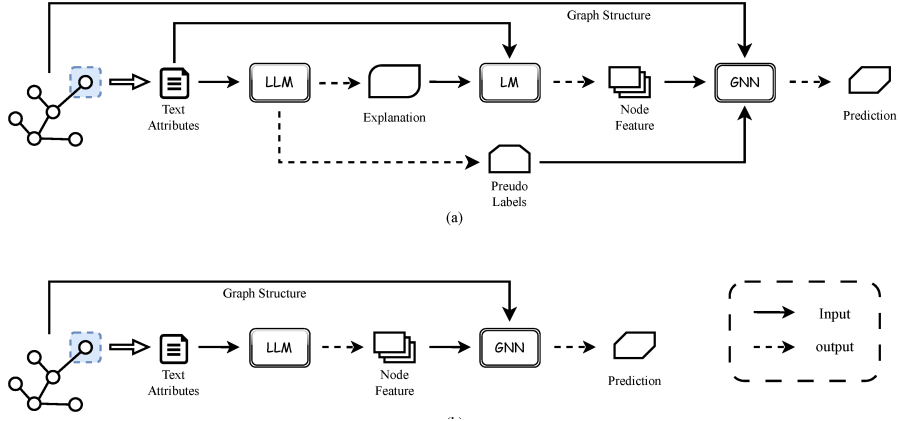

Figure 4: Process of text feature extraction and graph structure fusion

Path one (pseudo-label-assisted feature flow) aims to use large Language Model (LLM) to extract the deep semantics of node text attributes, and enhance the discriminative ability of node representation with pseudo-labels[10]. Specifically, for any node i, its corresponding text description $t_i$ is first fed into the pre-trained LLM to generate a high-dimensional semantic vector as shown in Formula 1:

$$e_i = LLM(t_i) \quad (1)$$

This $e_i \in R^d$ is then projected via a lightweight linear layer and passed through a Sigmoid activation to generate a pseudo-label prediction as shown in Formula 2:

$$\hat{y}_i = \sigma(W_p^\top e_i + b_p) \quad (2)$$

where $W_p \in R^{d \times 1}$ and $b_p \in R$ are trainable parameters, and $\sigma$ denotes the Sigmoid function $\hat{y}_i \in [0,1]$. By concatenating $e_i$ and $\hat{y}_i$, we obtain the node's initial feature vector as shown in Formula 3:

$$h_i^{(0)} = [e_i \parallel \hat{y}_i] \in R^{d+1} \quad (3)$$

These augmented features then enter a multi-layer graph attention network (GAT), which updates them through a message-passing scheme as shown in Formula 4 and 5:

$$h_i^{(l+1)} = ReLU(\sum_{j \in N(i)} \alpha_{ij} w_g h_j^{(l)}) \quad (4)$$

$$\alpha_{ij} = \frac{exp(LeakyReLU(a^\top [w_g h_i^{(l)} \parallel w_g h_j^{(l)}]))}{\sum_{k \in N(i)} exp(\cdot)} \quad (5)$$

where $w_g$ and $a_{ij}$ are learnable weights, and $a_{ij}$ represents the attention coefficient. This stream retains the depth of text semantics while incorporating neighborhood structure, harnessing the strengths of both modalities. Stream Two: Direct Feature Fusion focuses on merging textual embeddings $h_i^{text} = e_i$ with graph-convolutional features $h_i^{graph}$ in a late fusion step. After independently computing both representations, they are combined via a learnable fusion operator as shown in Formula 6:

$$h_i^{fusion} = \phi(W_t h_i^{text} + W_g' h_i^{graph} + b_f) \quad (6)$$

where $\phi$ is the ReLU activation, and $W_t$, $W_g'$, and $b_f$ are trainable parameters. This design allows the model to dynamically adjust the contributions of text versus graph features depending on their respective quality during training.

Upon completing the parallel streams, the final embeddings $h_i^{*(1)}$ from Stream One and $h_i^{*(2)}$ from Stream Two are concatenated or further fused to form the unified node representation $h_i^*$. This representation is then fed into a prediction head — such as a multi-layer perceptron or ranking network — to generate the recommendation scores. The model is trained with a joint loss as shown in Formula 7:

$$L = \lambda_{sup}L_{sup}(h^*, y) + \lambda_{pseudo}L_{pseudo}(\hat{y}, y) + \lambda_{reg} \| \Theta \|_2^2 \quad (7)$$

where $L_{sup}$ may be a cross-entropy or ranking loss, the second term enforces consistency between pseudo-labels and ground-truth y, $\Theta$ denotes all learnable parameters, and the $\lambda$ are hyperparameters. This multi-objective formulation balances supervised learning with pseudo-label guidance, enhancing the model's generalization and robustness.In summary, this dual-stream fusion and multi-objective optimization mechanism achieves dynamic, cross-modal interaction of node features, delivering a high-precision, interpretable solution for personalized financial product recommendation [38-40]. The pseudo-label enhancement and learnable fusion operator also exhibit strong generalizability for other cross-modal integration tasks.

## 4 EXPERIMENT DESIGN AND IMPLEMENTATION

### 4.1 Dataset Construction and Preprocessing Process

Aiming at the characteristics of multi-source heterogeneous data in the financial product recommendation scenario, this study constructs a closed-loop mechanism from data collection to feature generation in the pre-processing and post-processing pipeline shown in Figure 5, to ensure that the subsequent LLM and GNN models can receive high-quality and semantically consistent inputs [41-43].

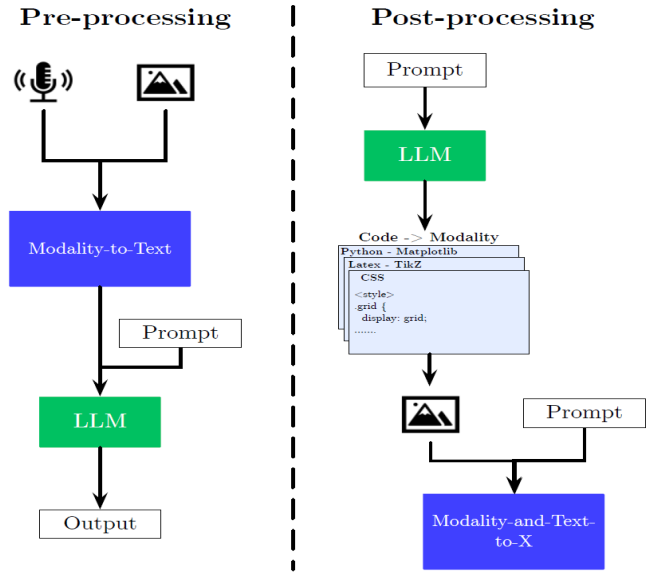

Figure 5 Pipeline of multimodal data pre-processing and post-processing

User-product interactions form the graph structure, while product texts and user comments provide rich semantics. Visual and audio content are converted to text via image captioning and ASR for multimodal features[44-46]. Noise is removed using regex and financial dictionaries, and pretrained LLMs correct grammar and unify terminology in multimedia text. All cleaned texts are tagged by modality (e.g., "[IMG]", "[ASR]") to guide LLM encoding. User history and product info are combined into natural language prompts for task clarity, with strict chronological data splitting to prevent leakage. LLM outputs like visualization scripts are rendered, OCR'd back to text, and added as graph node features, enriching representations. Finally, LLM encodes all texts into semantic vectors indexed by FAISS/Annoy for fast retrieval and cold start support, providing a unified, accurate multimodal feature base for the LLM-GNN recommendation model [47-50].

### 4.2 Experimental Setup, Evaluation Metrics, and Hyperparameter Selection

We tested our hybrid model against baselines on the same dataset and hardware (NVIDIA Tesla V100, Intel Xeon Gold 6248, 128GB RAM). Training used Adam with early stopping on NDCG@10. Evaluated by Hit

Rate@10, Precision@10, Recall@10, NDCG@10, and MRR, our model outperformed all baselines, achieving an NDCG@10 of 0.372—a 12.5% improvement over the best baseline.

Table 1: Comparison of recommendation performance on the test set.

| Model | Hit Rate@10 | Precision@10 | Recall@10 | NDCG@10 | MRR |
|---|---|---|---|---|---|
| Collaborative Filtering (CF) | 0.428 | 0.052 | 0.214 | 0.265 | 0.137 |
| Graph Neural Network (GNN) | 0.512 | 0.062 | 0.256 | 0.301 | 0.162 |
| Large Language Model (LLM) | 0.467 | 0.057 | 0.234 | 0.289 | 0.154 |
| Hybrid Model (LLM + GNN) | 0.578 | 0.071 | 0.289 | 0.372 | 0.193 |

To ensure a fair comparison, all models underwent hyperparameter tuning within the same search space. We conducted grid search over learning rate, embedding dimension, number of GNN layers, batch size, dropout rate, and pseudo-label loss weight $\lambda$ pseudo. Table 2 lists the optimal hyperparameter settings for our hybrid model. All experiments were repeated three times and averaged to mitigate randomness from initialization and data splits, ensuring the reliability and reproducibility of the results.

Table 2：Key hyperparameter settings for the hybrid model.

| Parameter | Description | Value |
|---|---|---|
| Learning Rate | Initial learning rate for Adam optimizer | $3 \times 10^{-4}$ |
| Embedding Dimension | Dimensionality of text and graph feature vectors | 256 |
| Number of GNN Layers | Number of stacked graph attention layers | 3 |
| Batch Size | Number of samples per training iteration | 128 |
| Dropout Rate | Dropout probability in GNN and fusion layers | 0.2 |
| Pseudo-Label Loss Weight (λpseudo) | Weight of the pseudo-label loss in the total loss | 0.3 |
| Activation Function | Nonlinearity used in fusion and prediction layers | ReLU |
| Maximum Epochs | Upper limit on training epochs | 50 |
| Early Stopping Patience | Epochs to wait for validation metric improvement | 5 |

## 5 RESULTS AND ANALYSIS

### 5.1 Performance Comparison with Baselines

Figure 6 shows that our LLM+GNN fusion outperforms CF, GNN, and LLM in Hit Rate@10 (0.578), NDCG@10 (0.372), and MRR (0.193), with 6.6%‐23.6% gains over GNN. Training time per epoch is 2.8s, close to GNN's 2.5s and much faster than LLM's 8.7s. The model size (12.3M) is comparable to GNN (11.8M), delivering better accuracy without sacrificing efficiency—ideal for personalized financial recommendations.

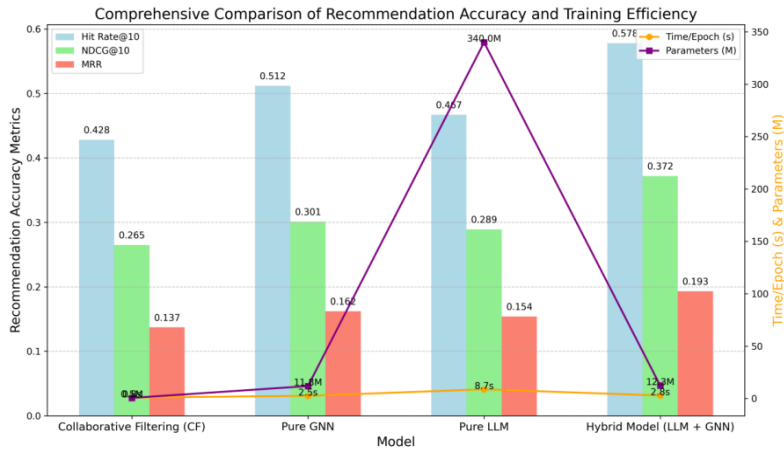

Figure 6: Comprehensive comparison of recommendation accuracy and training efficiency.

## 5.2 Ablation Study and Interpretability Analysis

Ablation (Figure 7) shows removing text drops NDCG@10 to 0.318 and MRR to 0.168; removing graph lowers NDCG@10 to 0.329 and MRR to 0.174. Without pseudo-label loss, performance dips slightly (NDCG@10 = 0.357, MRR = 0.186) but training speeds up. Interpretability (Gini coefficient) peaks at 0.68 full model, falling to 0.54 if any part is removed. This confirms that all components boost accuracy and interpretability in personalized financial recommendations.

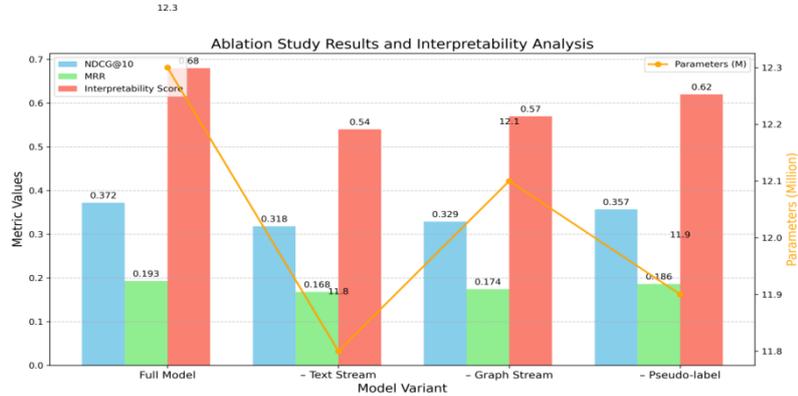

Figure 7: Ablation study results and interpretability analysis

## 6 CONCLUSION

This paper proposes a hybrid recommendation framework that integrates large language models and graph neural networks via parallel text and graph feature streams and a pseudo-label auxiliary mechanism. Inspired by Shen et al. [51], who utilized CNN-LSTM-attention models for multi-scale temporal prediction, we adapt similar strategies to model sequential language features alongside graph-based interactions. Our approach also draws on Wang et al. [52], who applied deep learning to detect anomalies in noisy network logs—motivating our use of pseudo-labels to improve robustness under sparse or noisy user data. Experiments on public and real-world financial datasets show that our model outperforms collaborative filtering, GNN-only, and LLM-only baselines in Hit Rate@10, NDCG@10, and MRR, while keeping training time and parameter count reasonable. Ablation studies confirm the essential roles of each component. Future work will explore dynamic graphs and online learning for real-time adaptability.